\documentstyle[preprint,aps,epsfig]{revtex}
\begin{document}
\tighten
\newcommand{\beq}{\begin{equation}}
\newcommand{\eeq}{\end{equation}}
\newcommand{\bea}{\begin{eqnarray}}
\newcommand{\eea}{\end{eqnarray}}
\newcommand{\ek}{\not{\varepsilon}}
\newcommand{\eek}{\vec{\varepsilon}}
\newcommand{\gp}{ {\cal G\/}_{F} }

\draft
\preprint{BETHLEHEM-Phys-HEP990215}
\title{
Hot Hypernuclear Matter in the Modified Quark Meson Coupling Model}
\date{February 22, 1999}

\author{
I. Zakout$^{1,2}$, H. R. Jaqaman$^{1}$\footnote{hjaqaman@bethlehem.edu}, 
S. Pal$^{2}$, H. St\"ocker$^{2}$ and W. Greiner$^{2}$}
\address{$^1$ Department of Physics, Bethlehem University,
P.O. Box 9, Bethlehem, Palestine}
\address{$^2$ Institut f\"ur Theoretische Physik, 
J. W. Goethe Universit\"at, Robert Mayer Sta$\beta$e 8-10,
Postfach 11 1932, D-60054, Frankfurt am Main, Germany}  

\maketitle
\begin{abstract}
Hot hypernuclear matter is investigated in an explicit $SU(3)$ quark  
model based on a mean field description of nonoverlapping baryon bags 
bound by the self-consistent exchange of scalar $\sigma, \zeta$ 
and vector $\omega, \phi$ mesons. 
The $\sigma, \omega$ mean fields are assumed to couple to the $u,d$-quarks
while the $\zeta, \phi$ mean fields are coupled to the $s$-quark. 
The coupling constants of the mean fields with the quarks 
are assumed to satisfy $SU(6)$ symmetry. 
The calculations take into account the medium dependence 
of the bag parameter on the scalar fields $\sigma, \zeta$.
We consider only the octet baryons $N,\Lambda,\Sigma,\Xi$ 
in hypernuclear matter.
An ideal gas of  the strange mesons $K$ and  $K^{*}$ 
is introduced to keep zero net strangeness density. 
Our results for symmetric hypernuclear matter show that a phase transition 
takes place at a critical temperature around $180$ MeV in which the scalar 
mean fields $\sigma, \zeta$ take nonzero values at zero baryon density.
Furthermore, the bag contants of the baryons decrease significantly at
and above this critical temperature indicating the onset of quark
deconfinement. The present results imply that the onset of quark
deconfinement in $SU(3)$ hypernuclear matter is much stronger than 
in $SU(2)$ nuclear matter. 
\end{abstract}

\narrowtext
\section{introduction}

The majority of nuclear phenomena have been successfully described  
in relativistic mean-field theory using only hadronic degrees of 
freedom\cite{Walecka75,Serot86,Serot92}. 
However, due to the observations which revealed the medium modification 
of the internal structure of the baryons\cite{EMC}, it has become essential 
to explicitly incorporate the quark-gluon degrees of freedom while 
respecting the established model based on the hadronic degrees of freedom 
in nuclei. 
One of the first models put forward along these lines is the quark-meson
coupling (QMC) model, proposed by Guichon\cite{Guichon}, 
which describes nuclear matter as a collection of non-overlapping MIT bags
interacting through the self-consistent exchange of scalar $\sigma$ and
vector $\omega$ mesons in the mean field approximation with the meson
fields directly coupled to the quarks\cite{Blunden,Saito94}. 
The scalar $\sigma $ meson is supposed to simulate the exchange 
of correlated pairs of pions and may
represent a very broad resonance observed in $\pi\pi$ scattering, while
the vector $\omega$ meson is identified with the actual meson having a mass
of about 780 MeV. 
In the chiral models, the scalar $\sigma$ and vetcor $\omega$ mean fields 
represent the $u,d$ quark condensates\cite{Papa98,Papa99}. 
The QMC model thus incorporates explicitly the quark degrees of freedom 
and this has nontrivial consequences\cite{Blunden,Saito94}.
It also have been extended to study superheavy finite
nuclei\cite{Tsushima,Saito97}.

In the so-called modified quark meson coupling model (MQMC)\cite{Jin,Jin2}, 
it has been further suggested that including a medium-dependent bag parameter 
may be essential for the success of relativistic nuclear phenomenology. 
It was found that when the bag parameter is significantly reduced in the
nuclear medium with respect to its free-space value, large cancelling
isoscalar Lorentz scalar and vector potentials for the nucleon 
in nuclear matter emerge naturally. 
Such potentials are comparable to those suggested by relativistic nuclear
phenomenology and finite density QCD sum rules\cite{Cohen}. 
The density-dependence of the bag parameter is introduced by coupling it to
the scalar meson field\cite{Jin}. 
This coupling is motivated by invoking the nontopological soliton model 
for the nucleon\cite{soliton}. 
In this model a scalar soliton field provides the confinement of the quarks. 
This effect of the soliton field is, roughly speaking, mimicked by 
the introduction of the bag parameter
in the Bag Model. When a nucleon soliton is placed in a nuclear environment,
the scalar soliton field interacts with the scalar mean field. 
It is thus reasonable to couple the bag parameter 
to the scalar mean fields\cite{Jin,Jin2}.

The QMC model was extended to finite temperatures 
to investigate the liquid-gas phase transition in nuclear matter\cite{Song}. 
Recently, the MQMC model has been also extended to finite 
temperature\cite{Panda,Jaqaman,Jaqamanb} 
and has been applied to the study of the properties of nuclear matter where 
it was found that the bag parameter decreases appreciably above a critical
temperature $T_c\approx 200$ MeV indicating the onset of quark
deconfinement\cite{Jaqaman}. 
The effect of  glueball exchange as well
as a realization of the broken scale invariance of quantum chromodynamics
has also been investigated in the MQMC model through the introduction 
of a dilaton field\cite{Jaqamanb}. 
It was found that the introduction of the dilaton potential improves 
the shape of the saturation curve at T=0 and affects hot 
nuclear matter significantly\cite{Jaqamanb}.

In the  present work, we extend the MQMC model to hot hypernuclear matter 
by introducing the scalar $\zeta$ and vector $\phi$  mean fields that are
coupled to the $s$-quark in addition to the $\sigma$ and $\omega$ fields 
which are coupled to the $u,d$-quarks.  
The $\zeta$ and $\phi$ fields are identified with the real mesons 
having masses $m_{\zeta}=975$ and $m_{\phi}=1020$ MeV respectively
\cite{Papa98,Papa99,Schaffner}.  
Hypernuclear matter is considered to contain the octet 
$p,n,\Lambda,\Sigma^+,\Sigma^0,\Sigma^-,\Xi^0$ and $\Xi^-$ 
baryons\cite{Mueller98}. 
We introduce an ideal gas of kaons to keep
zero net strangeness density $\rho_S=0$.  
We simplify the calculations by considering
symmetric hypernuclear matter whereby the octet baryons
reduce to 4 species: $2N, \Lambda, 3\Sigma, 2\Xi$. 

The outline of the paper is as follows. In section II, we present the MQMC
model for hypernuclear matter at finite temperature, together with the
details of the self-consistency conditions for the vector and scalar mean
fields. In section III, we discuss our results and present our conclusions.


\section{The Quark Meson Coupling Model for Hypernuclear Matter}

The quark field $\psi_{q}(\vec{r},t)$ inside a bag of radius $R_i$  
representing  a baryon of species $i$ satisfies the Dirac equation 
\begin{eqnarray}
\left[ i\gamma^{\mu}\partial_{\mu}- m_{q}^{0}
+(g_{\sigma}^{q}\sigma -g_{\omega}^{q}\omega_{\mu}\gamma^{\mu})\delta_{qr}
+(g_{\zeta}^{q}\zeta-g_{\phi}^{q}\phi_{\mu}\gamma^{\mu})\delta_{qs}
\right]\psi_{q}(\vec{r},t)=0.
\label{Dirac}           
\end{eqnarray}
where $m_{q}^{0}$ is the current mass of the quarks of flavor $q$ and
where $r$ refers to the up or down quarks and $s$ refers to the
strange quark.
The Kronecker deltas insure that the u, d quarks are coupled only to the
$\sigma$ and $\omega$ fields  while the s-quark is coupled only to the
$\zeta$ and  $\phi$ fields. In the mean field approximation all the meson
fields are treated classically and, for nuclear matter in this
approximation, these fields are translationally invariant. 
Moreover, because of rotational invariance,  the space-like components 
of the vector fields  vanish  so that  we have 
$\omega_{\mu}\gamma^{\mu}=<\omega_0>\gamma^{0}=\omega\gamma^{0}$ 
and 
$\phi_{\mu}\gamma^{\mu}=<\phi_0>\gamma^{0}=\phi\gamma^{0}$. 

The single-particle quark and antiquark
energies in units of $R_i^{-1}$ for quark flavor $q$ are given by 
\begin{eqnarray}
{\epsilon_q}^{n\kappa}_\pm=\Omega_q^{n\kappa}
\pm \left(   g_{\omega}^{q}\omega R_i\delta_{qr}
+ g_{\phi}^{q}\phi R_i\delta_{qs} \right)
\label{EPN}             
\end{eqnarray}
where 
\begin{eqnarray}
\Omega_q^{n\kappa}=\sqrt{ {x^q_{n\kappa}}^{2} 
+ R_i^{2}{m^{*}}^{2}_{q} } 
\label{Omegnk}          
\end{eqnarray}
and 
\begin{eqnarray}
m^{*}_{q}=m^{0}_{q}-g^{q}_{\sigma}\sigma\delta_{qr}
-g^{q}_{\zeta}\zeta\delta_{qs},
\label{effmass}         
\end{eqnarray} 
are the effective quark kinetic energy and effective quark mass, 
respectively. 
The boundary condition for each quark of flavor $q$  at the bag surface 
is given by 
\begin{eqnarray}
i\gamma\cdot \hat{n} \psi_{q}^{n\kappa}({x^q_{n\kappa}})
=\psi_{q}^{n\kappa}({x^q_{n\kappa}}),
\label{roots}           
\end{eqnarray}
which determines the quark momentum $x^q_{n\kappa}$ in the state
characterized by specific values of $n$ and $\kappa$. 
For a given value of  the
bag radius $R_i$ for baryon species $i$ and the scalar fields 
$\sigma$ and $\zeta$, the quark momentum 
$x^q_{n\kappa}$ is  determined by the boundary condition 
Eq. (\ref{roots}) which, 
for quarks of flavor $q$ in a spherical bag, reduces to 
$j_0({x^{q }_{n\kappa}})=\beta_{q} j_1({x^{q}_{n\kappa}})$, where 
\begin{eqnarray}
\beta_{q}=
\sqrt{ \frac{
{\Omega^{*}}_{q}^{n\kappa}(\sigma,\zeta)
-R_i m^{*}_{q}(\sigma,\zeta) }
{{\Omega^{*}}_{q}^{n\kappa}(\sigma,\zeta)
+R_i m^{*}_{q}(\sigma,\zeta) }}.
\label{beta}            
\end{eqnarray}  

The quark chemical potential $\mu_{q}$, 
assuming that there are three quarks in the baryon bag, 
is determined from $3 =\sum_{q}n_q$ where $n_q$ is the
number of quarks of flavor q and is determined
by\cite{Panda,Jaqaman,Jaqamanb}
\begin{eqnarray}
 n_q ={\sum_{n\kappa}}\left[
 \frac{1}{e^{({\epsilon_q}^{n\kappa}_{+}/R-\mu_{q})/T}+1}
-\frac{1}{e^{({\epsilon_q}^{n\kappa}_{-}/R+\mu_{q})/T}+1}
\right].
\label{nq}            
\end{eqnarray}
The total energy from the quarks  and antiquarks of each baryon 
of species $i$ is 
\begin{eqnarray}
E^i_{tot}=\sum_{n\kappa q}
 n_{q}\frac{\Omega_{q}^{n\kappa}}{R}\left[
 \frac{1}{e^{({\epsilon_{q}}^{n\kappa}_{+}/R_i-\mu_{q})/T}+1}
+\frac{1}{e^{({\epsilon_{q}}^{n\kappa}_{-}/R_i+\mu_{q})/T}+1}  
\right].
\label{Etot}         
\end{eqnarray}
The bag energy  for baryon species $i$ is given by 
\begin{eqnarray}
E^i_{bag}=E^i_{tot} - \frac{Z_i}{R_i}
+\frac{4\pi}{3}R_i^{3}  B_i(\sigma,\zeta).
\label{Ebag}         
\end{eqnarray}
where $B_i=B_i(\sigma,\zeta)$ is the bag parameter. In the simple QMC model,
the bag parameter $B$ is taken as $B_{0}$ corresponding to its value for a
free baryon. The medium effects are taken into account in the MQMC 
model \cite{Jin,Jin2}
by coupling the bag parameter to the scalar meson fields. 
In the present work we generalize the coupling suggested
in the latter references to  the case of two  scalar meson fields 
by using  the following ansatz for the bag parameter
\begin{eqnarray}
B_i=B_{0}\exp\left[-\frac{4}{3}
\frac{((n_{u}+n_{d})g^{B}_{\sigma}\sigma
+n_{s}g^{B}_{\zeta}\zeta)}{ M_{i}}\right]
\label{bagCon}          
\end{eqnarray}
with $g^{B}_{\sigma}$ and $g^{B}_{\zeta}$ as additional parameters.
The spurious center-of-mass momentum in the bag is subtracted to obtain 
the effective baryon mass\cite{Fleck} 
\begin{eqnarray}
M^{*}_{i}=\sqrt{{E^i_{bag}}^2 - <{p^{2}_{cm}}>^i},
\label{MNSTAR}          
\end{eqnarray}
where 
\begin{eqnarray}
<{p^{2}_{cm}}>^i=\frac{<x^{2}>^{i}}{R_i^{2}}
\label{PCM}             
\end{eqnarray}
and 
\begin{eqnarray}
<x^{2}>^{i}= \sum_{n\kappa q} n_{q}{x^{q}_{n\kappa}}^{2}
\left[
 \frac{1}{e^{({\epsilon_{q}}^{n\kappa}_{+}/R_i-\mu_{q})/T}+1}
+\frac{1}{e^{({\epsilon_{q}}^{n\kappa}_{-}/R_i+\mu_{q})/T}+1}
\right].
\label{xseq}            
\end{eqnarray}
The bag radius $R_{i}$ for baryon species $i$ is obtained through the 
minimization of the baryon mass with respect to the bag 
radius\cite{Saito94,Jin,Panda,Jaqaman,Jaqamanb}
\begin{eqnarray}
\frac{\partial M^{*}_{i}}{\partial R_{i}}=0.
\label{MNR}             
\end{eqnarray}
The total energy density at finite temperature $T$ and at finite baryon
density $\rho_{B}$ reads 
\begin{eqnarray}
\varepsilon&=&
\sum^{Baryons}_i \frac{\gamma_i}{(2\pi)^{3}} \int
d^{3}k\sqrt{k^{2}+{M_{i}^{*}}^{2}}(f_{i}+\overline{f}_{i})
+\frac{1}{2}m^{2}_{\omega} \omega^2
+\frac{1}{2}m^{2}_{\phi} \phi^2
+\frac{1}{2}m^{2}_{\sigma}\sigma^{2}
+\frac{1}{2}m^{2}_{\zeta}\zeta^{2} 
\nonumber \\
&+&   \varepsilon^{id}_{K},
\label{Edensity}        
\end{eqnarray}
where $\gamma_i$ is the spin-isospin degeneracy factor of baryon 
species $i$ and  where the last term corresponds to the energy 
density of the $K$-mesons treated here as an ideal gas. 
In Eq. (\ref{Edensity}) $f_{i}$ and $\overline{f}_{i}$ are 
the Fermi-Dirac distribution functions for the baryons 
and antibaryons of species $i$,
\begin{eqnarray}
f_{i}=\frac{1}{e^{(\epsilon_i^{*}-\mu^{*}_{i})/T}+1},
\label{fB}             
\end{eqnarray}
and 
\begin{eqnarray}
\overline{f}_{i}=\frac{1}{e^{(\epsilon_i^{*}+\mu^{*}_{i})/T}+1},
\label{fBAR}           
\end{eqnarray}
where $\epsilon_i^{*}$ and $\mu^{*}_{i}$ are, respectively, the effective  
energy and effective chemical potential of baryon species $i$. 
These are given by $\epsilon_i^{*}=\sqrt{ k^{2}+{M^{*}_{i}}^{2} }$
and  
$\mu^{*}_{i}=
B_i\mu_B+S_i\mu_S-\left(g_{\omega i}\omega+g_{\phi i}\phi \right)$ 
where $B_i$ and $S_i$ are the baryon and strangeness quantum numbers 
and where $g_{\omega i}=(n_{u}+n_{d})g^{q}_{\omega i}$ 
and $g_{\phi i}=n_{s}g^{q}_{\phi i}$ .
The chemical potentials $\mu_B, \mu_S$ are determined by 
the self-consistency equations for the total baryonic density 
\begin{eqnarray}
\rho_{B}=\frac{1}{(2\pi)^{3}}
\sum_i B_i \gamma_i \int d^{3}k(f_{i}-\overline{f}_{i}),
\label{rhoB1}         
\end{eqnarray}
and the total strangeness density
\begin{eqnarray}
\rho_{S}=\frac{1}{(2\pi)^{3}}
\sum^{Baryons}_i S_i \gamma_i
\int d^{3}k(f_{i}-\overline{f}_{i})
+\sum^{Kaons}_i S_i {\rho^{id}_{Ki}} = 0
\label{rhoS1}         
\end{eqnarray}
where in the last equation we have introduced the contribution 
of an ideal gas of $K$ and $K^{*}$ mesons to make the total 
strangeness density vanish identically. 
The vector mean fields are determined  by 
\begin{eqnarray}
\omega=\sum_i \frac{g_{\omega i}}{m^{2}_{\omega}}B_i\rho_{i},
\label{vecomeg}        
\end{eqnarray}
and
\begin{eqnarray}
\phi=\sum_i \frac{g_{\phi i}}{ m^{2}_{\phi}}B_i\rho_{i}.
\label{vecphi}         
\end{eqnarray}

The pressure is the negative of the grand thermodynamic potential density
and is given by 
\begin{eqnarray}
P&=&
\frac{1}{3}\sum_i\frac{\gamma_i}{(2\pi)^{3}}\int d^{3} k
\frac{k^{2}}{\epsilon_i^{*}}(f_{i}+\overline{f}_{i})
+\frac{1}{2}m^{2}_{\omega}\omega^{2}
+\frac{1}{2}m^{2}_{\phi}\phi^{2}
-\frac{1}{2}m^{2}_{\sigma}\sigma^{2}
-\frac{1}{2}m^{2}_{\zeta}\zeta^{2}
+P^{id}_{K},
\label{pressurD}   
\end{eqnarray}
where the summation $i$ runs over the 8 species of the baryon octet which 
reduces for symmetric hypernuclear matter to 4 species with 
$\gamma_i=4,2,6$ and $4$ for $N$, $\Lambda$, $\Sigma$ and $\Xi$, 
respectively. 
In Eq. (\ref{pressurD}) $P^{id}_{K}$ is the pressure of the ideal gas 
of $K$ mesons. 
The density of the $K$-mesons of species $i$ is given by 
\begin{eqnarray}
{\rho^{id}_{K}}_i=\frac{\gamma^K_i}{(2\pi)^{3}}
\int d^{3}k (b_i - \overline{b}_i)
\label{Mesdens}       
\end{eqnarray}
where the spin-isospin degeneracy $\gamma^K_i=4,6$ for $K, K^{*}$, 
respectively.
The Bose-Einstein distribution functions for the $K$ mesons are given by 
\begin{eqnarray}
b_i=\frac{1}{e^{[\sqrt{k^2+M^2_i}-\mu_i]/T}-1}
\label{boson}         
\end{eqnarray}
and 
\begin{eqnarray}
\overline{b}_i=\frac{1}{e^{[\sqrt{k^2+M^2_i}+\mu_i]/T}-1}
\label{bosonbar}      
\end{eqnarray}
where $\mu_i=S_i\mu_S$ is the chemical potential of $K$-meson of species $i$. 
The total energy density and total 
pressure of the $K$-meson ideal gas are given by 
\begin{eqnarray}
\varepsilon^{id}_{K}=
\sum_i \frac{\gamma^K_i}{(2\pi)^{3}}
\int d^{3}k \sqrt{k^{2}+M^{2}_i}
(b_i-\overline{b}_i),
\label{EdensK}       
\end{eqnarray}
and 
\begin{eqnarray}
P^{id}_{K}=\frac{1}{3}
\sum_i \frac{\gamma^K_i}{(2\pi)^{3}}
\int d^{3}k 
\frac{k^2}{\sqrt{k^{2}+M^{2}_i}}
(b_i-\overline{b}_i),
\label{PdensK}       
\end{eqnarray}
respectively.

The scalar mean fields $\sigma$  and $\zeta$ are  determined through 
the minimization of the
thermodynamic potential or the maximizing of the pressure  with respect to 
these fields.  The pressure depends explicitly on the
scalar mean fields  through the $\frac{1}{2}m_{\sigma}^2\sigma^2$ and 
$\frac{1}{2}m_{\zeta}^2\zeta^2$ terms in Eq. (\ref{pressurD}). 
It also depends on the baryon effective masses $M^{*}_{i}$ which in turn 
also depend on $\sigma$ and $\zeta$. 
If we write the pressure
as a function of $M_{i}^{*}$, $\sigma$ 
and $\zeta$\cite{Jaqaman,Jaqamanb}, 
the extremization of $P(M^{*}_{i},\sigma, \zeta)$ with respect to the
scalar mean field $\sigma$ can be written as 
\begin{eqnarray}
\left( \frac{\partial P}{\partial \sigma} \right)_{\zeta}=
\sum_i
\left( \frac{\partial P}{\partial M^{*}_{i}} \right)_{\mu_{B},T}
\left( \frac{\partial M^{*}_{i}}{\partial \sigma} \right)_{\zeta}
+\left(\frac{\partial P}{\partial \sigma}\right)_{\{M^{*}_{i}\}}=0,
\label{dP1}         
\end{eqnarray}
where 
\begin{eqnarray}
\left(\frac{\partial P}{\partial \sigma}\right)_{\{M^{*}_{i}\}}=
- m^{2}_{\sigma}\sigma,
\label{dPdseg}      
\end{eqnarray}
with a similar experssion for the extremization of
$P(M^{*}_{i},\sigma, \zeta)$ with respect 
to the scalar mean field $\zeta$.
The derivative of the pressure with respect to effective 
mass $M^{*}_{i}$ reads 
\begin{eqnarray}
\left( \frac{\partial P}{\partial M^{*}_{i}} \right)_{\mu_{i},T}=
&-&\frac{1}{3}\frac{\gamma_i}{(2\pi)^{3}}
\int d^{3} k
\frac{k^{2}}{{\epsilon^{*}}^{2}}\frac{M^{*}_{i}}{\epsilon_i^{*}}
\left[f_{i}+\overline{f}_{i}\right]
\nonumber \\
&-&\frac{1}{3}\frac{\gamma_i}{(2\pi)^{3}}
\frac{1}{T}\int d^{3} k
\frac{k^{2}}{\epsilon_i^{*}}\frac{M^{*}_{i}}{\epsilon_i^{*}}
\left[f_{i}(1-f_{i})+\overline{f}_{i}(1-\overline{f}_{i})\right]
\nonumber \\
&-&\frac{1}{3}\frac{\gamma_i}{(2\pi)^{3}}
\frac{1}{T}g_{\omega i}
\left(\frac{\partial \omega}{\partial M^{*}_{i}}\right)_{\mu_{i},T}
\int d^{3} k
\frac{k^{2}}{\epsilon_i^{*}}
\left[f_{i}(1-f_{i})-\overline{f}_{i}(1-\overline{f}_{i})\right]
\nonumber \\
&-&\frac{1}{3}\frac{\gamma_i}{(2\pi)^{3}}
\frac{1}{T}g_{\phi i}
\left(\frac{\partial \phi}{\partial M^{*}_{i}}\right)_{\mu_{i},T}
\int d^{3} k
\frac{k^{2}}{\epsilon_i^{*}}
\left[f_{i}(1-f_{i})-\overline{f}_{i}(1-\overline{f}_{i})\right]
\nonumber \\  
&+& m^{2}_{\omega} \omega 
\left(\frac{\partial \omega}{\partial M^{*}_{i}}\right)_{\mu_{i},T}
\nonumber \\
&+&m^{2}_{\phi} \phi
\left(\frac{\partial \phi}{\partial M^{*}_{i}}\right)_{\mu_{i},T}.
\label{dPdMef}    
\end{eqnarray}
Since the baryon chemical potential $\mu_{B}$ and temperature are treated as
input parameters, the variation of the vector mean field $\omega$ with
respect to the effective baryon mass $M^{*}_{i}$ at a given value of the
baryon density $\rho_{B}$ reads 
\begin{eqnarray}
\left(\frac{\partial \omega}{\partial M_{i}^{*}}\right)_{\mu_{i},T}=
-\frac{
\left[g_{\omega i} / m^{2}_{\omega}\right]
\left[\gamma_i/(2\pi)^{3}\right]\left[1/T\right]\int d^{3}k
\frac{M_{i}^{*}}{\epsilon^{*}_i}
\left[f_{i}(1-f_{i})-\overline{f}_{i}(1-\overline{f}_{i})\right]}
{1+\sum_j \left[ g^{2}_{\omega j}/ m^{2}_{\omega} \right]
\left[\gamma_j/ (2\pi)^{3}\right]
\left[1 /T \right] \int d^{3}k
\left[f_{j}(1-f_{j})+\overline{f}_{j}(1-\overline{f}_{j})\right]}.
\label{OME}        
\end{eqnarray}
with a similar expression for 
$\left({\partial \phi}/{\partial M_{i}^{*}}\right)_{\mu_{i},T}$.
The coupling of the scalar mean fields $\sigma$ and $\zeta$ with the quarks 
in the non-overlapping MIT bags through the solution of the point like 
Dirac equation should satisfy the self-consistency condition. 
These constraints are essential to obtain the correct solution of
the scalar mean fields $\sigma$ and $\zeta$.

\section{Results and Discussions}

We have studied hypernuclear matter at finite temperature
using the modified quark meson coupling model 
which takes the medium-dependence of the bag into account. 
We choose a direct coupling  of the bag parameter 
to the scalar mean fields $\sigma$ and $\zeta$ in the form given 
in Eq. (\ref{bagCon}).
The bag  parameter is taken as that adopted by Jin and Jennings\cite{Jin} 
 $B^{1/4}_{0}=188.1$ MeV and the free nucleon bag radius 
$R_{0}=0.60$ fm.  We have taken the current quark masses to be  
$m_{u}=m_{d}=0$ and $m_{s}= 150$ MeV.
For $g_{\sigma}^{q}=1$, the values of the vector meson coupling constant
and the parameter $g_{\sigma}^{B}$, as fitted from the saturation 
properties of nuclear  matter, are given 
as $g^{2}_{\omega}/4\pi=(3g^q_{\omega})^2/4\pi =5.24$ 
and 
${g^{B}_{\sigma}}^{2}/4\pi=({3g^B_\sigma}^q)^{2}/4\pi$=3.69.
The $Z_i=2.03, 1.814, 1.629$ and $1.505$ are chosen to reproduce the
baryon masses at their experimental values
$M_{N,\Lambda,\Sigma,\Xi}=$939, 1157, 1193, and 1313 MeV
respectively. Normal nuclear matter saturation density 
is taken as ${\rho_B}_0=0.17$ fm$^{-3}$.
The extra  coupling constants needed to couple the
scalar and vector mean fields $\zeta$ and $\phi$ to 
the $s$-quark are chosen to satisfy  $SU(6)$ symmetry
where $|g_\zeta^q|=\sqrt{2} |g_\sigma^q|$ and 
$|g_\phi^q|=\sqrt{2} |g_\omega^q|$ and 
$|{g_\zeta^B}^q|=\sqrt{2} |{g_\sigma^B}^q|$.
If it is assumed  that the mean fields $\zeta$ and $\phi$ 
are positive definite, then all the coupling constants are
positive and the absolute value signs become redundant.

The $\sigma$ mean field is supposed to simulate 
the exchange of correlated pairs of pions and may represent
a very broad resonances observed in $\pi \pi$
scattering and fixed at $m_\sigma=550$ MeV,
while the vector $\omega$ meson is identified with the actual meson 
whose mass is $m_\omega=783$ MeV.
Since the mean fields, $\sigma$ and $\omega$, are considered as 
$<u\overline{d}>$ condensates, they interact only with
$u,d$-quark in the baryons.
On the other hand, the scalar and vector mean fields
$\zeta, \phi$ are considered as actual mesons with 
$m_\zeta=975$ MeV and $m_\phi=1020$ MeV, respectively.
They are considered as $<s\overline{s}>$ condensates
and interact only with the $s$-quarks in the baryons.
This picture is consistent with the chiral models\cite{Papa98,Papa99}.
We fixed the total strangeness density $\rho_S$ to  $0$
by introducing  an ideal gas of $K$ and $ K^*$ mesons 
where $m_K=$ 495 MeV and  $m_{K^*}=$ 892 MeV. 
The contribution of other $K$ mesons was found to be negligible.
It is supposed in the ideal gas limit that the Kaons 
do not  interact with the mean fields $\zeta$ and $\phi$. 
The extension to the case that the Kaons interact with 
the $\zeta$ and $\phi$ fields will be considered in a future work.

We first solve Eqs. (\ref{rhoB1}), (\ref{rhoS1}), (\ref{vecomeg}) and 
(\ref{vecphi}) self consistently
for given values of temperature $T$ and densities $\rho_B$ and $\rho_S$
to determine the baryonic  and strangeness  chemical potentials $\mu_B$ and 
$\mu_S$, respectively. 
These constraints are given in terms of the effective baryon 
masses $M^*_i$ which depend on the bag radii $R_i$,
the quark chemical potentials $\mu^i_q$ and the mean fields. 
For given values of the scalar fields $\sigma, \zeta$ and vector
fields $\omega, \phi$, 
the quark chemical potential and bag radius of species $i$ 
are obtained using the self consistency conditions 
Eqs. (\ref{nq}) and (\ref{MNR}), respectively.
The pressure is evaluated for specific values of temperature $T$
and chemical potentials $(\mu_B, \mu_S)$ which now become input 
parameters. We then determine the values of $\sigma$ and $\zeta$
by using the extermization conditions as given in Eq. (\ref{dP1}).
These constraints take into account the coupling of the quark with 
the scalar mean fields in the frame of the point like Dirac equation 
exactly\cite{Saito94,Jin}. 

The dependence of the baryon effective masses $M_{N,\Lambda,\Sigma,\Xi}^{*}$ 
on the  total baryonic density $\rho_B$ and temperature is shown 
in Fig. 1 where 
it is seen that the baryon masses decrease with baryonic density except 
at the highest temperatures where the effective masses become 
almost density-independent.
Moreover, for a given  baryonic density $\rho_{B}$  it is seen that 
as the temperature is increased the mass $M_{i}^{*}$ of species $i$
first increases slightly up to about $T=150$ MeV and then decreases 
rather rapidly for $T>150$ MeV. 
These results are displayed in a different manner in Fig. 2 which plots 
the baryon masses $M^{*}_i$ as a function of  $T$ for 
$\rho_B=0$ fm $^{-3}$ and ${\rho_B}_0=0.17$ fm $^{-3}$.
It is seen that the effective baryon masses 
$M_{N,\Lambda,\Sigma,\Xi}^{*}$ 
with ${\rho_B}_0=0.17$ fm $^{-3}$ are less than those
with zero baryonic density. Moreover, 
the effective baryonic  masses
$M_{N,\Lambda,\Sigma,\Xi}^{*}$ increase only slightly, if at all,  with
temperature up to about $T= 150$ MeV beyond  which they decrease rapidly.  
This behaviour is qualitatively similar to our earlier results for normal 
nuclear matter\cite{Furnstahl,Jaqaman,Jaqamanb} where the rapid decrease 
in the nucleon's effective mass was, however, 
found to start at rather higher temperatures $T>200$ MeV.
This rapid decrease of $M_{i}^{*}$ with increasing temperature
resembles a phase transition at high temperatures and low density, when
the system becomes a dilute gas of baryons in a sea of
baryon-antibaryon pairs\cite{Furnstahl}.

In Fig. 3, we display the scalar mean fields $\sigma$ and $\zeta$  
as  functions of the total baryonic density $\rho_B$ for various 
temperatures.
It is seen  that the value of $\sigma$ initially  decreases
with increasing temperature for temperatures less than 150 MeV.
The scalar mean field $\zeta$ is almost negligible for such
low temperatures. 
However, as the temperature reaches 150 MeV there are 
indications of an increase in $\sigma$ at low baryon densities
where it  attains a nonzero value at $\rho_{B}=0$.
For still higher temperatures, the situation is more dramatic with 
the value of $\sigma$  increasing with temperature for all values of 
$\rho_{B}$. This is also qualitatively similar to our earlier results  for  
$SU(2)$ nuclear matter\cite{Jaqaman,Jaqamanb}.
Furthermore,  the scalar mean field $\zeta$ becomes 
important for $T>150$ MeV and also increases with  temperature 
for all values of $\rho_{B}$. 
An  interesting new feature here is that the scalar mean fields $\sigma$
and $\zeta$ tend to take almost constant values irrespective 
of density at high  temperatures  which was not seen in our earlier 
calculations for $SU(2)$ nuclear matter even at temperatures as high  
$240$ MeV .
Fig. 4 displays $\sigma$ and $\zeta$ versus $T$ for
$\rho_B=0$  and ${\rho_B}_0=0.17$ fm $^{-3}$. 
It is seen that the $\sigma$ field has a nonzero (and almost constant)   
value only at the higher density until the 
temperature reaches about $T=150$ MeV when a rapid increase  sets in 
at both densities. 
The increase at  zero density is actually more dramatic and 
the $\sigma$ field rapidly attains values equal to those occuring 
at ${\rho_B}_0=0.17$ fm $^{-3}$.
This is another indication of a phase transition to a system 
of baryon-antibaryon pairs. 
The behaviour of the $\zeta$ field is qualitatively similar  except that
its value is negligible at both densities for temperatures less than 
about  $T=150$ MeV. 

In Fig. 5, we display the baryonic density dependence 
of the bag parameters 
for $N,\Lambda,\Sigma,\Xi$ for different values of the temperature.
For each baryon, the  bag parameter 
increases  with temperature 
for temperatures less than 150 MeV.
However, the situation  is completely reversed after the phase transition 
takes place. For temperatures  $T>150$ MeV
the bag parameters display a dramatic 
decrease with temperature for all densities. This can be seen more clearly in 
Fig. 6 which displays $B_i$ vs $T$ for
$\rho_B=0$ fm $^{-3}$ and ${\rho_B}_0=0.17$ fm $^{-3}$.
The bag parameters are almost constant until the 
temperature exceeds $T=150$ MeV when they start to  decrease rapidly.
This indicates the onset of quark deconfinement above the critical 
temperature: 
at high enough temperature and/or baryon density there is a phase transition 
from the baryon-meson phase to the  quark-gluon phase. This behaviour 
is also in qualitative agreement with our earlier results for $SU(2)$ 
nuclear matter\cite{Jaqaman} except that the decrease here is
more dramatic indicating that the phase transition is much stronger
than in ordinary nuclear matter. 
Our results are also comparable to those obtained from lattice QCD calculations 
which have so far only explored the zero baryon density 
axis of the phase diagram in a meaningful way. 
The  lattice  results with 2 light quark flavors, 
indicate that the transition from hadronic matter to a quark-gluon plasma 
occurs 
at a temperature $T\approx 140$  MeV and that for low densities it may not
be a phase transition  at all but what is called a rapid crossover\cite{PRLcross}. 

Fig. 7 displays the relative abundance $\rho_i/\rho_B$  for each baryon species.
At low temperatures the nucleons  $N$ are almost the only constituents.
However, the contribution of the hyperons starts to be noticeable 
when the temperature reaches $100$ MeV and becomes more  important when the 
temperature is increased to $T>150$ MeV. 
At temperatures $T>200$ MeV the contribution of the nucleons falls down to about 
half the total baryonic density. 
This can also be seen in Fig. 8 where we display 
the ratio of the baryonic strangeness density to the total baryonic density. 
As mentioned earlier, we keep the net strangeness of the system fixed to zero by 
introducing Kaons so that $\rho_S=\rho^{Baryons}_S+\rho^{Mesons}_S=0$. 
It is seen that the net strangeness of the baryons $\rho^{Baryons}_S$
is  small at low temperatures $T<150$ MeV (note the logarithmic scale).
However, as the temperature increases the net strangeness of the
baryon octet becomes significant.
For $T>150$ MeV, the ratio 
$\rho^{Baryons}_S/\rho_B$ becomes of order one. 
Therefore, the $K$-meson plays a significant
role at high temperatures and probably also in the phase transition. 

In Fig. 9, we display the relative abundances of the various baryon species 
$\rho_i/\rho_B$ versus temperature at normal nuclear matter density
${\rho_B}_0=0.17$ fm $^{-3}$.
It is seen that the  the nucleons are dominant at  low temperatures  and their  
relative abundance $\rho_N/\rho_B$  
decreases very slowly at first and deviates very little from $1$ until 
the temperature reaches about 100 MeV when  $\rho_N/\rho_B$  
starts to decrease rapidly and becomes  negligibly small for temperatures larger 
than about  280 MeV. 
On the other hand,  the abundance of the hyperons is negligible at low temperatures 
and  increases significantly as the
temperature is increased beyond $T=100$ MeV. 
They become more abundant than the nucleons for temperatures larger than 200 MeV. 
The relative abundance of hyperons in high energy heavy ion collisions can therefore 
be used as a simple thermometer to measure the temperature of the hot nuclear matter 
produced in the reaction and probably to study the phase transition to the quark gluon plasma. 

In conclusion, we have generalized the MQMC model to the case of hot hypernuclear matter 
by introducing two new meson fields that couple to the strange quarks and using  
$SU(6)$ symmetry to relate the coupling constants. 
The results are qualitatively  similar to those obtained for $SU(2)$ nuclear 
matter including the onset of  quark deconfinement. 
It is observed, however,  that in  this model quark deconfinement in $SU(3)$ 
(or hypernuclear) matter is much stronger than in $SU(2)$  (or  normal)
nuclear matter and that hyperons become more abundant than nucleons at high 
temperatures $T>200$ MeV. 
An ideal gas of Kaons was introduced to keep zero net strangeness density. 
The interaction of the Kaons with the meson fields will be considered in a future work. 

\acknowledgments
Financial support by  the Deutsche Forschungsgemeinschaft through the  
grant GR 243/51-1 is gratefully acknowledged.


\clearpage

\begin{figure}
\centerline{\mbox{\epsfig{file=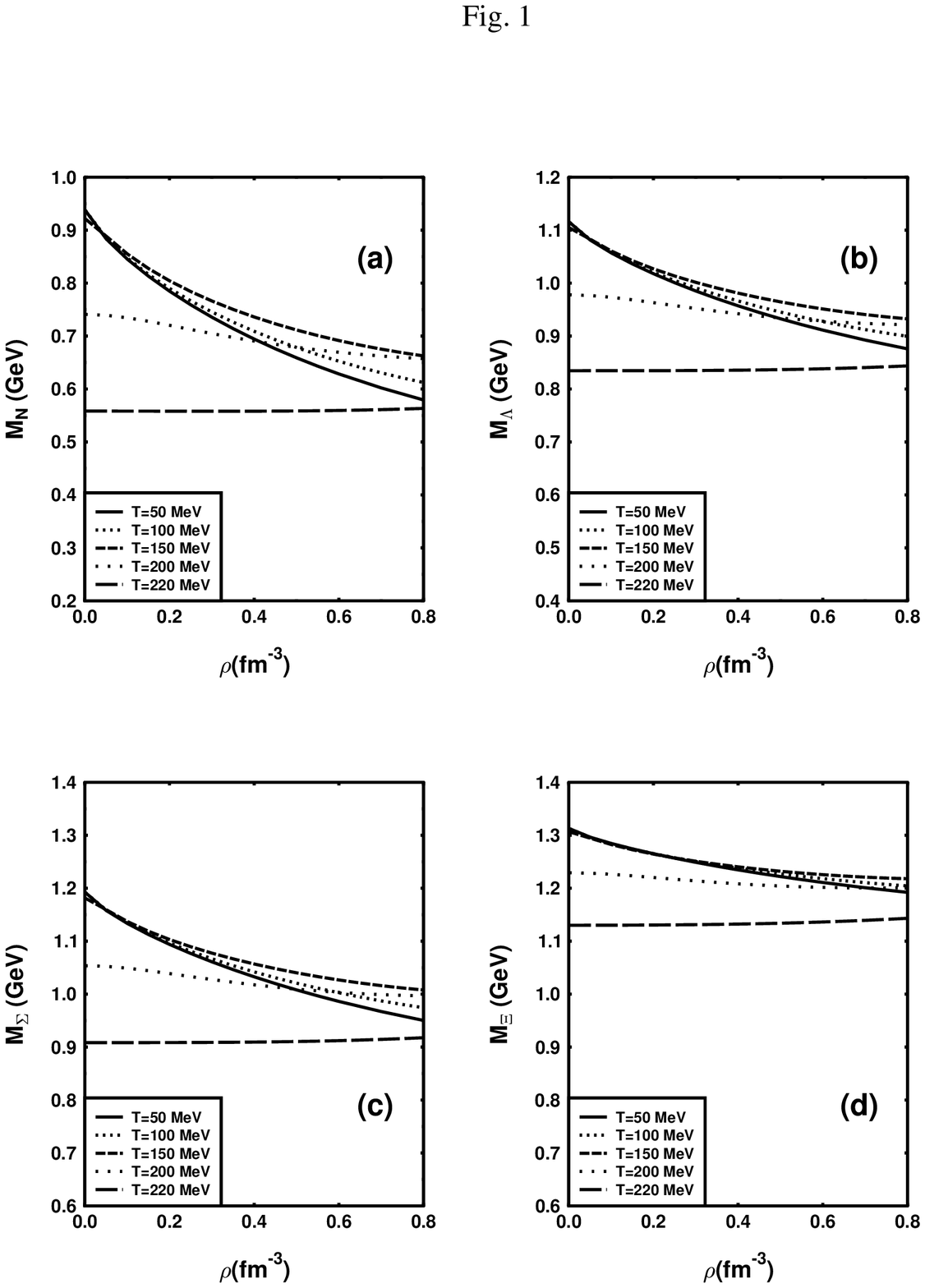,width=0.8\linewidth}}}
\vspace{1truein}
\caption{
The effective baryon masses $M^{*}_i$ in hypernuclear matter as a function
of the total baryonic density $\rho_B$ for different values of temperature.}
\label{fa1}
\end{figure}

\begin{figure}
\centerline{\mbox{\epsfig{file=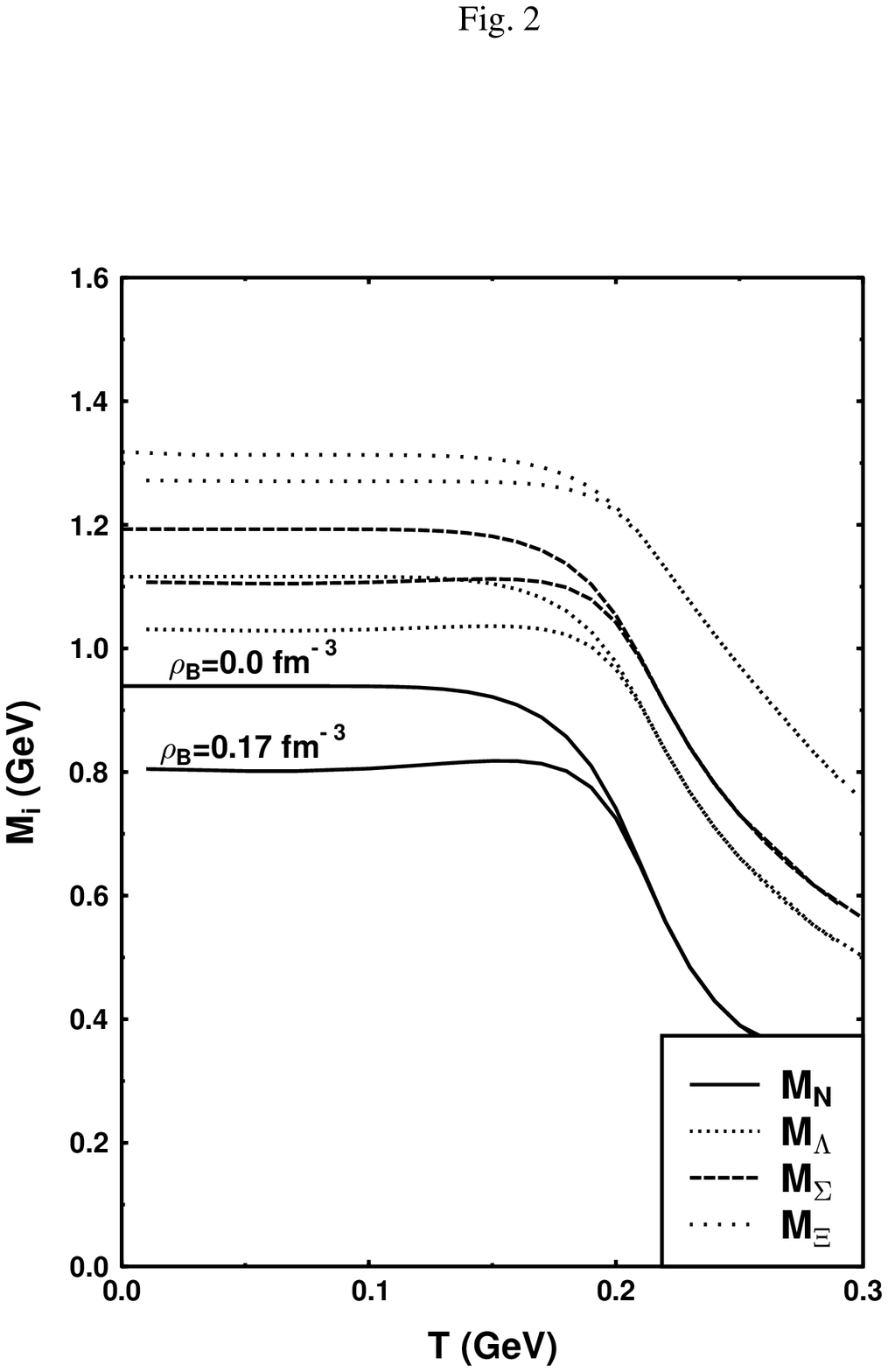,width=0.8\linewidth}}}
\vspace{1truein}
\caption{
The effective baryon masses $M^{*}_i$ in hypernuclear matter as a function  
of temperature for $\rho_B=0$ and $\rho_B=0.17$ fm $^{-3}$.}
\label{fa2} 
\end{figure}
 
\begin{figure}
\centerline{\mbox{\epsfig{file=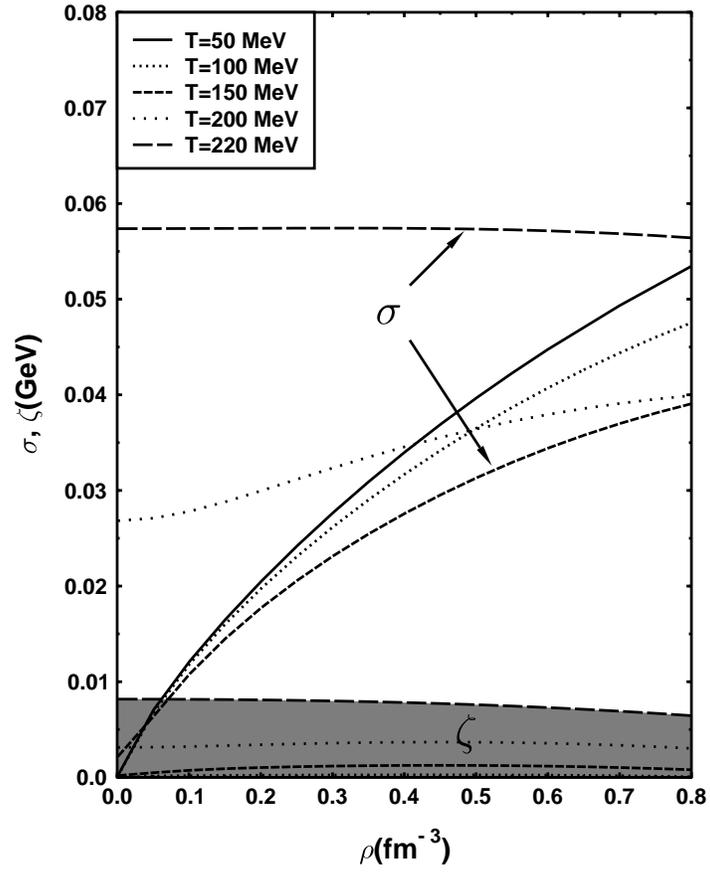,width=0.8\linewidth}}}
\vspace{1truein}
\caption{
The mean scalar fields $\sigma, \zeta$ as a function of the total baryonic 
density $\rho_B$ for different values of temperatures.}
\label{fa3}
\end{figure}                                                        

\begin{figure}
\centerline{\mbox{\epsfig{file=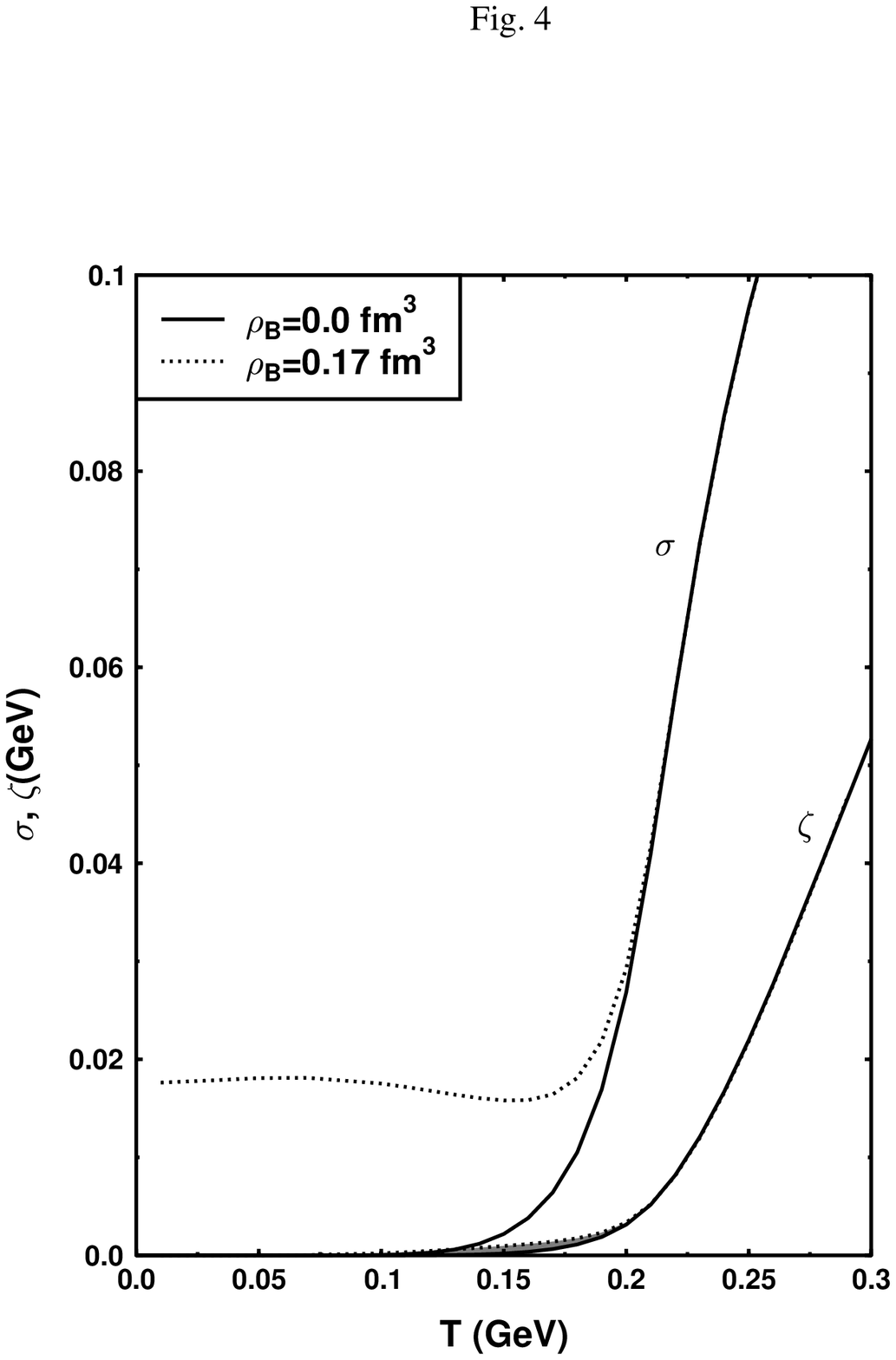,width=0.8\linewidth}}}
\vspace{1truein}
\caption{
The mean scalar fields $\sigma, \zeta$ versus the 
temperature for  $\rho_B=0$ and $\rho_B=0.17$ fm $^{-3}$.}
\label{fa4}
\end{figure}                                                   

\begin{figure}
\centerline{\mbox{\epsfig{file=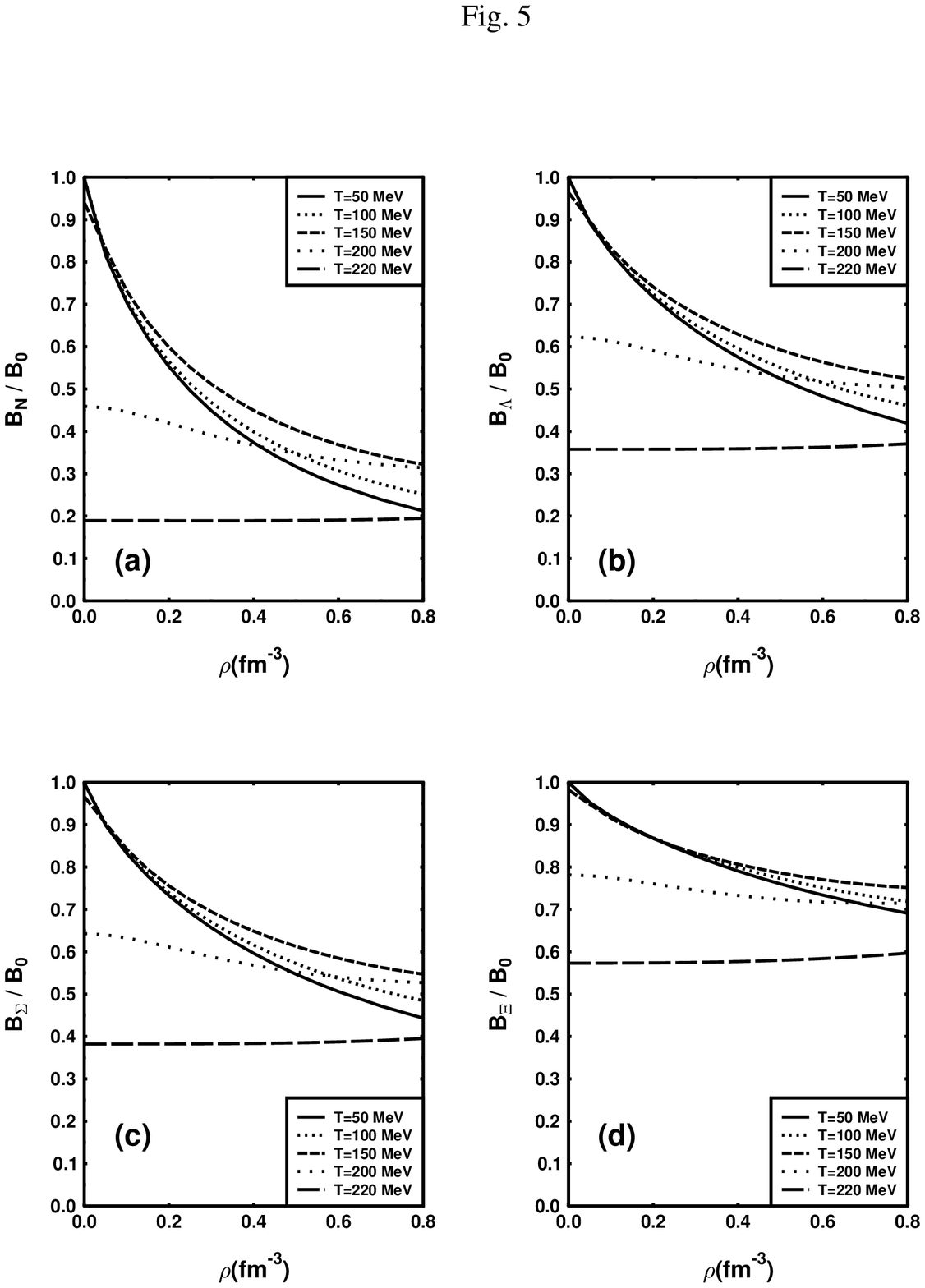,width=0.8\linewidth}}}
\vspace{1truein}
\caption{
The bag parameters $B_i$ of the nucleon and hyperons
versus the total baryonic density $\rho_B$ for different values 
of temperature.}
\label{fa5}
\end{figure}

\begin{figure}
\centerline{\mbox{\epsfig{file=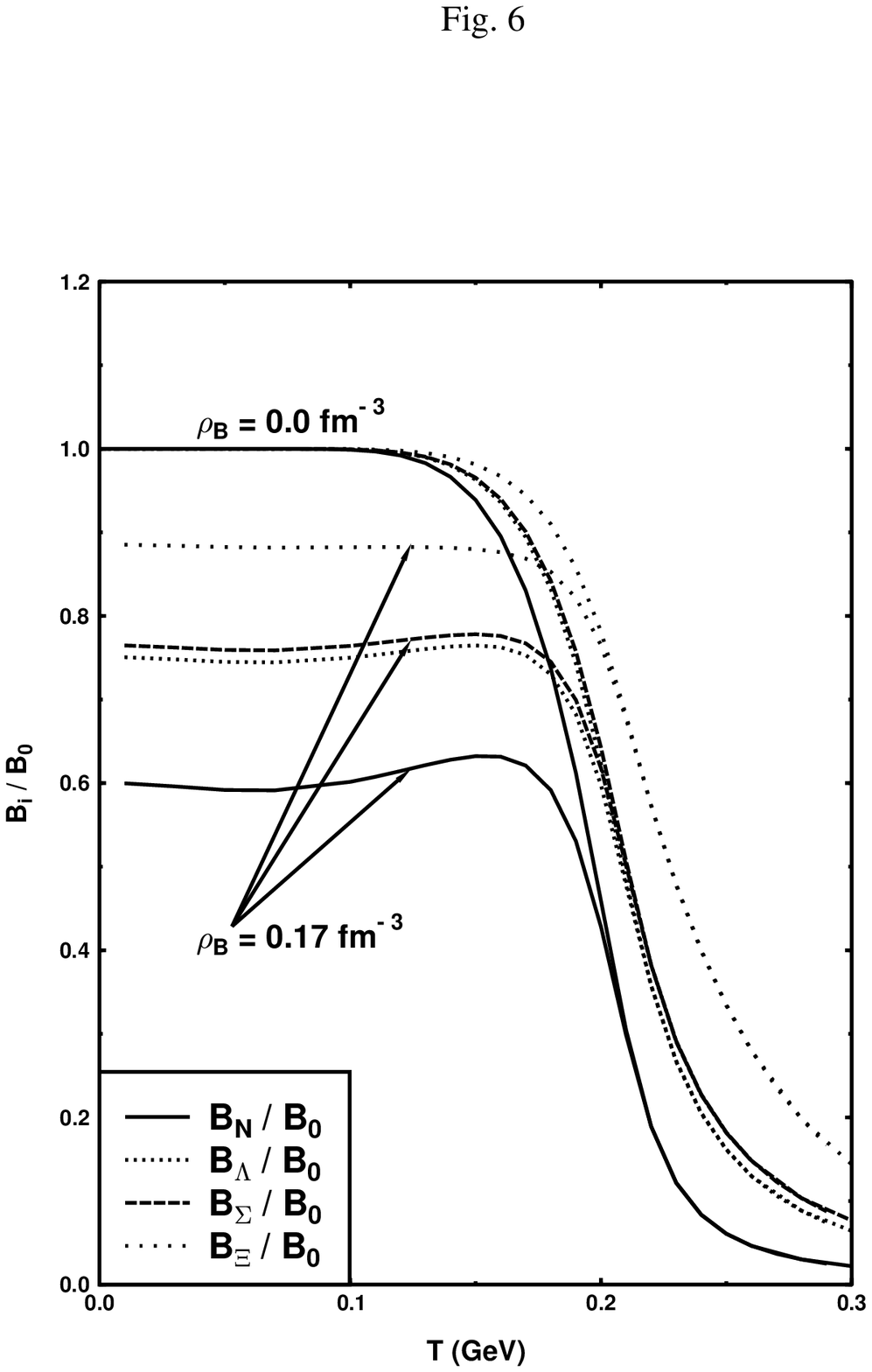,width=0.8\linewidth}}}
\vspace{1truein}
\caption{
The bag parameters $B_i$ of the nucleon and hyperons
versus the temperature $T$
for  $\rho_B=0$ and $\rho_B=0.17$ fm $^{-3}$.}
\label{fa6}
\end{figure}               

\begin{figure}
\centerline{\mbox{\epsfig{file=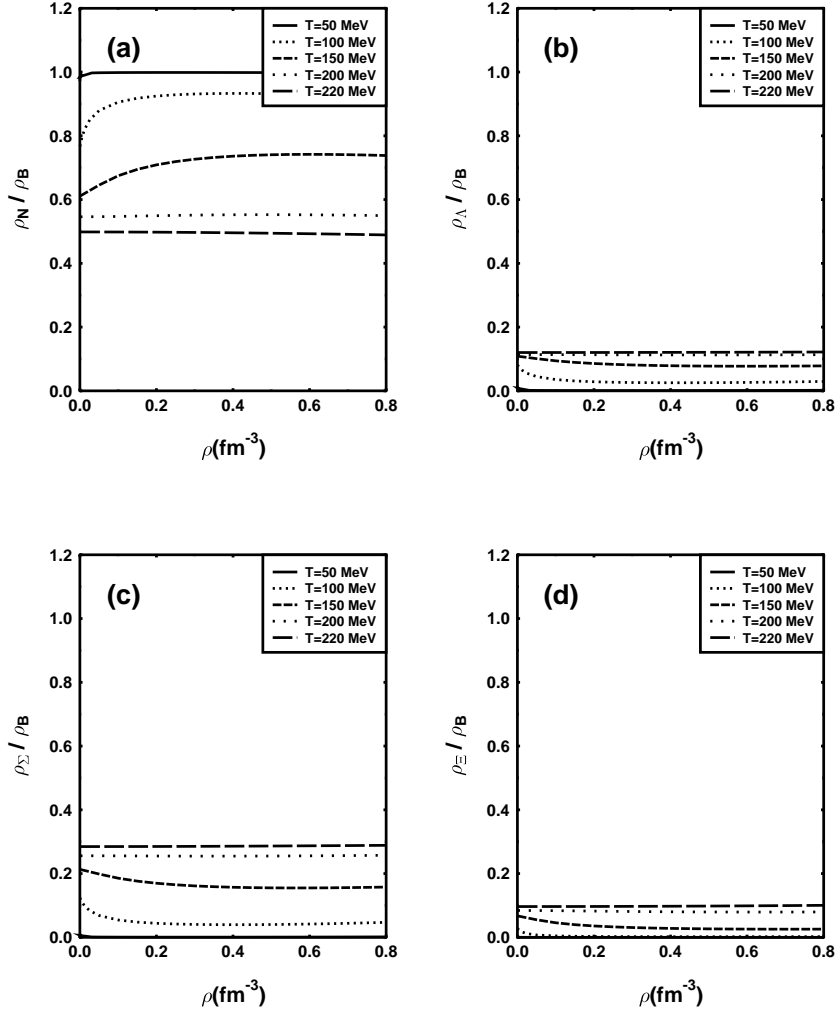,width=0.8\linewidth}}}
\vspace{1truein}
\caption{The relative abundance of  baryon species $i$, 
$\frac{\rho_i}{\rho_B}$,
versus the total baryonic density $\rho_B$ 
for different temperature.}
\label{fa7}
\end{figure}

\begin{figure}
\centerline{\mbox{\epsfig{file=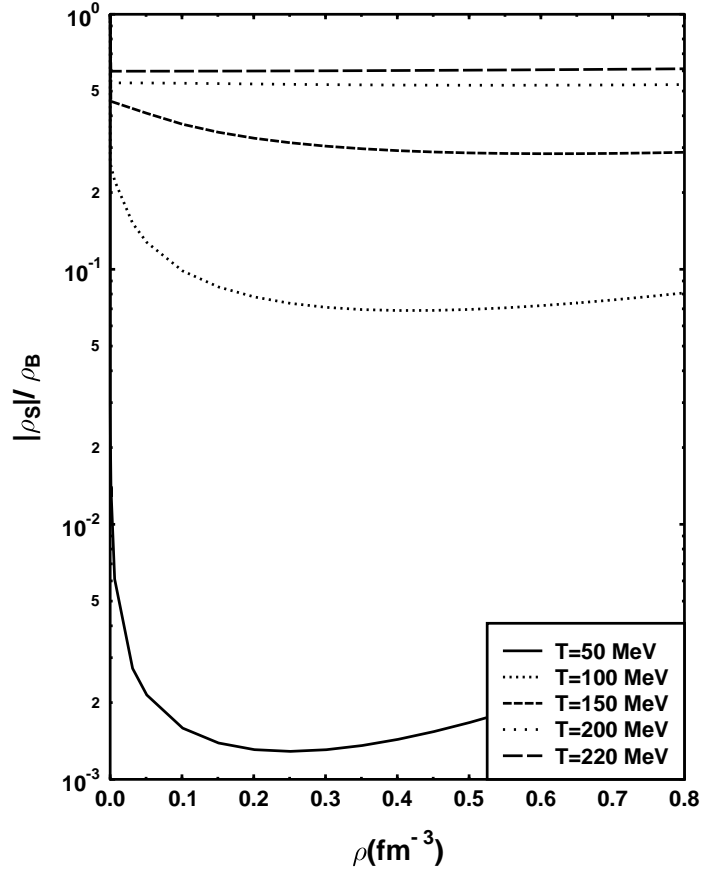,width=0.8\linewidth}}}
\vspace{1truein}
\caption{The absolute value of the ratio of the baryonic stangeness density
to the total baryonic density $\rho_S/\rho_B$ 
 as a function of the total baryonic density $\rho_B$ for various temperatures.}
\label{fa8}
\end{figure}

\begin{figure}
\centerline{\mbox{\epsfig{file=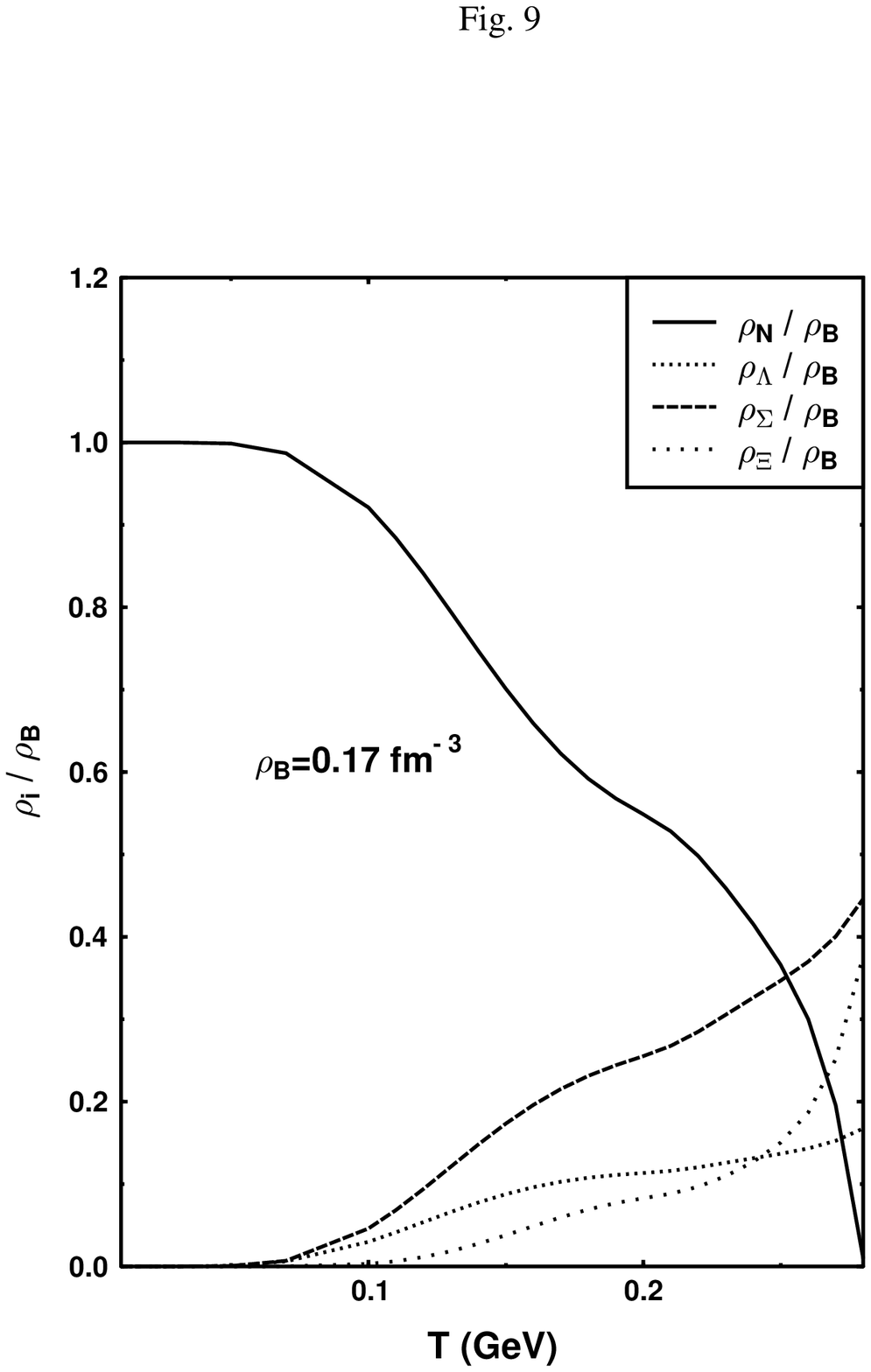,width=0.8\linewidth}}}
\vspace{1truein}
\caption{The relative abundance of baryon species $i$, 
$\frac{\rho_i}{\rho_B}$, versus temperature $T$ 
for $\rho_B=0.17$ fm $^{-3}$.}
\label{fa9}
\end{figure}

\end{document}